\documentstyle[a4,cite,12pt,aps,epsfig]{revtex}

\def\nnd{\end{document}}

\def\nn{\nonumber}
\def\nnb{\nonumber}

\def\be{\begin{equation}}
\def\ee{\end{equation}}
\def\mn{\mu\nu}
\newcommand{\bea}{\begin{eqnarray}}
\newcommand{\eea}{\end{eqnarray}}

\def\wbra#1{\left  ( #1  \right ) }
\def\sbra#1{\Big   ( #1  \Big   ) }
\def\mbra#1{\bigg  [ #1  \bigg  ] }
\def\bbra#1{\Bigg \{ #1  \Bigg \} }
\def\wpng#1#2{\frac{1}{#2} \wbra{#1}}

\def\mpng#1#2{\frac{1}{#2} \mbra{#1}}
\def\bpng#1#2{\frac{1}{#2} \bbra{#1}}
\def\fwpng#1#2#3{\frac{#1}{#3} \wbra{#2}}

\def\cma{\,,}

\def\brk{\nnb \\}
\def\jsep{\nnb \\ &&}
\def\wsep{ \nnb \\ &&}
\def\ssep{\right. \nnb\\ && \left.}
\def\eed{\end{document}}

\def\al{\alpha}

\def\al{\alpha}
\def\be{\beta}

\def\de{\delta}
\def\al{{\alpha}}
\def\my#1{ #1 }

\def\nkf#1{ #1 }

\def\mbkf#1{\big [ #1 \big ]}

\def\sbbkf#1{\bigg ( #1 \bigg )}

\def\bbbkf#1{\bigg \{ #1 \bigg \}}

\makeatother

\begin{document}

\draft

\title{The radiative corrections 
of the electroweak chiral Lagrangian 
with renormalization group equations}

\author{
             Qi-Shu YAN\footnote{
        E-mail Address: yanqs@post.kek.jp}\\
	Theory Group, KEK,  Tsukuba, 	 305-0801, Japan\\
}
\bigskip

\address{\hfill{}}

\maketitle


\begin{abstract}
We show the computational procedure of 
the renormalization of 
the electroweak chiral Lagrangian (the 4D Higgsless
standard model), and provide one simplified
version of its one-loop renormalization group equations, which
we demonstrate its simplicity and reliability.
By analyzing the solutions of the one-loop renormalization 
group equations of the electroweak chiral Lagrangian, we
study the parameter space of the precision test parameters 
at ultraviolet cutoff with the current low energy
experimental constraints. 
We find that the region of the permitted parameter space 
can be greatly amplified (1 to 2 order) by the 
radiative corrections of
those undetermined anomalous couplings. 
\end{abstract}


The effective Lagrangian method is a bottom-to-up 
and model-independent approach to understand 
experimental data \cite{wein, georgi, eff}. 
The electroweak chiral Lagrangian (EWCL) 
method (it was also called
as the non-linear gauged sigma model or
4D Higgsless standard model in some references) 
\cite{long, appel, bernard} is an effective
field method which is expected to describe the electroweak
physics without a Higgs in the 
standard model, since till now there
is no direct evidence for the existence of such a 
scalar field. In principle, the EWCL with low energy 
fermions can explain all the experiment data \cite{ewcl-fit} 
below the energy scale $\mu=200$ GeV or so. We can 
also use this theoretical
framework to understand physics beyond $\mu=200$ GeV or so
before we will find 
new particles and new resonances in the future experiments.
However, from the fact that the
unitarity of the scattering amplitude of
the longitudinal vector bosons (according to the equivalent
theorem \cite{ewet}, this corresponds to the Goldstone scatterings) 
will be violated, we can deduce that new physics must be below
$4 \, \pi \, v$, a few TeV \cite{scattering}.

Before a new resonance is found at experiments,
with its anomalous couplings as parameters the EWCL 
can describe all possible effects of the new physics
to the electroweak bosonic sector,
either the strong couplings and weak couplings 
origin of electroweak symmetry breaking, or
other possible scenarios (like extra dimensional
Higgsless model \cite{higgsless}). 
There is a vast number of literature about the 
phenomenology of the EWCL \cite{phen-ewcl} (For a review
please refer \cite{review}).

In this article, using the dimensional
regularization and modified minimal subtraction scheme, we 
show the main steps of computational procedure of the
renormalization of the EWCL and provide a simplified version of
renormalization group equations (RGEs) which is easy to use. 
We also study the radiative corrections of anomalous couplings 
by solving the RGEs. We analyze the effects of anomalous couplings
to the running of weak coupling constant $g$, 
the $U(1)$ coupling $g'$, the vacuum expectation
value $v$, and the quadratic anomalous couplings.

Traditional wisdom might regard that the radiative corrections
of anomalous couplings should be very small and there is no
need to consider them. The argument is
that in QCD (B physics, for instance), it is due to large
$g_s$ at low energy region that makes 
the QCD radiative corrections important. But in the electroweak
case, from a few TeV down to $100$ GeV or so, the weak coupling
constant $g_W$ is much smaller than $g_s$. So the electroweak
running effects should be small. 
Furthermore, the traditional power counting from the 
naive dimensional analysis
\cite{manahar} told us that the anomalous operators belong 
to $O(p^4)$, so the radiative corrections of them should 
belong to $O(p^6)$, therefore they should be of two loops 
contributions of $O(p^2)$ operators and are supposed to be
tiny.

However, we would like to point out several 
facts for the EWCL that this wisdom has
neglected. 1) The naive dimensional 
analysis only counts the dimension of the
operators and only qualitatively estimate the importance of
the operators, but in a realistic case, 
the importance of a certain 
operator depends not only on its momentum dimension power 
but also on the magnitude of its effective couplings. 
For instance, although ${\cal L}_0$ in the EWCL has momentum power $O(p^2)$,
due to the smallness of its coupling $\beta$, normally
we know its effect to physics processes would be less
important as ${\cal L}_{WZ}$, of which the effective coupling
is of $O(1)$.
2) As matter of fact, some of $\al_i$ might be of order $O(1)$: 
2.a) this phenomenological assumption is 
not contradicted to the current experiments \cite{smewfit,bagger,lepexp,lep}; 
2.b) In principle , the strong electroweak theories \cite{strong} 
do not forbidden that some of anomalous 
couplings are $O(1)$ (Of course, 
in the naive technicolor model \cite{ntm}, all anomalous
couplings are of the same order and can not be of $O(1)$).
Some of $\al_i$ might reach $O(1)$, 
as in non-standard Higgs model \cite{nonstandard}, for instance,
since $\al_i$ can receive large tree level contributions.
Therefore, those operators in $O(p^4)$ with 
large couplings should be more important than or at least
equal to those operators in $O(p^2)$ with tiny couplings.
3) In order to keep the generality and universality of the
effective field theory method, before we really know the
underlying theories, we treat all 
the relevant and marginal operators as of the same importance.
This phenomenological assumption is
necessary as it is complimentary to the pure 
theoretical argument and estimate.

However, in order to control and estimate the contributions of
the higher loop and higher order operators,
we modify the power counting rule by setting the momentum power
of $\al_i$, the couplings 
in the anomalous operators, to be $-2$.
We will find that the radiative corrections
of $\al_i$ always appear in the combination as $\al_i g_w^2$, $\al_i {g'}^2$,
and $\al_i G^2$, as shown in the RGEs 
in Eq. (\ref{rge0}---\ref{rge1}). These terms have the momentum power $O(p^0)$,
which is the same as that of the contribution of Goldstones 
from the $O(p^2)$ operators. This modified power counting rule
has been introduced when we considered the $SU(2)$ chiral Lagrangian \cite{su2}
and works quite well.

Generally speaking, if the anomalous couplings are
of $O(1)$, the unitarity of amplitudes of some processes
of the theory might be violated, $W_L W_L \rightarrow W_L W_L$ for instance. 
This imposes the correlation between the magnitude of 
anomalous couplings and ultraviolet cutoff $\Lambda$: {\it i.e.} the 
higher the ultraviolet cutoff, the smaller the magnitude 
of the anomalous couplings, if the validity of the effective description should
be preserved.

Loop calculations in the EWCL 
have been started quite a long time ago, and
it is well known that the quartic, quadratic, and logarithmic 
divergences \cite{div} are witnessed. Reference \cite{hagiwara} 
introduces a Higgs field as a regulator. While in 
the reference \cite{bijj}, the authors used the method 
of higher covariant derivatives ($O(p^6)$) 
to regulate and parameterize the quartic 
and quadratic divergences, and the cutoffs of different types of
vector bosons are set to be different. However, in the 
framework of the effective theory method \cite{georgi}, with the 
${\overline {MS}}$ renormalization scheme and dimensional 
regularization, we can handle these divergences consistently 
and systematically. And after regularizing the
quartic, quadratic and logarithmic divergence by using the
dimensional regularization, anomalous couplings are logarithmically 
dependent on the universal cutoff.

Several works have analyzed the radiative correction
to the precision test parameters \cite{burgess,bijj,wud}
of anomalous operators by assuming the large anomalous
couplings. 
The formula obtained there are quite complicated. 
Here we provide the RGE method, which is also not  
simple. However, for the physics permitted region,
we can keep only the linear terms involving
the anomalous couplings and the RGE method become very simple.
Furthermore, we find that this simplified version of 
RGEs is quite reliable for the reasonable parameter 
space of the EWCL.

There are several technical difficulties in 
the actual calculation procedure: 
1) There is a mixing between 
the photon and Z bosons mixing in the kinetic sector;
2) The number of relevant vertices is large, and how to 
construct the counter terms is a task; 
3) Compared with the SU(2) chiral Lagrangian, 
the calculation is complicated due to the non-degenerate 
masses of A, Z and W bosons (and Goldstone bosons 
$\xi_Z$ and $\xi^{\pm}$).
To overcome these technical difficulties, we use the
path integral method \cite{path}, Stueckelberg transformation \cite{stuckelberg},
background field method \cite{bfm}, heat kernel method \cite{hk} 
and short-distance expansion method \cite{sdem} to extract the divergences.
These methods work quite well, as we have demonstrated in the
$SU(2)$ case \cite{su2}. Below we outline the main 
computational steps to the renormalization of EWCL and
RGEs (Please refer to our long paper for details \cite{longppr}).  

\indent {\bf Step 1:} The EWCL \cite{long,bernard,appel} can be formulated as
\bea
{\cal L}_{EW} &=&{\cal L}_{EW}^{p^2} + {\cal L}_{EW}^{p^4} +\cdots\label{ewla}\\
{\cal L}_{EW}^{p^2}&=&{\cal L}_B\cma\\
{\cal L}_{EW}^{p^4}&=&\beta {\bar {\cal L}_0} +\sum_{i=1}^{10} {\alpha_i } {\bar {\cal L}_i}\,
\eea
where
\bea
{\cal L}_{B}&=&- {\bar H_1} -  {\bar H_2} + {\bar {\cal L}_{WZ}}\cma\\
{\bar H_1}&=& {1\over 4 g^2}  W_{\mn}^a W^{a \mn} \label{Lh1}\cma\\
{\bar H_2}&=& {1\over 4 g'^2} B_{\mn}   B^{\mn}\label{Lh2}\cma\\
{\bar {\cal L}_{WZ}}&=&{v^2 \over 4} tr(V \cdot V)\,,
\eea
After using the relations of Lie algebra and
the classic equation of motion to eliminate
the redundant operators, the complete
Lagrangian ${\cal L}_{EW}^{p^4}$ without violating discrete
$C$, $P$, and $CP$ symmetries includes the
following independent operators \cite{long, appel}:
\bea
\label{ewclb}
{\bar {\cal L}_0}&=&  {v^2 \over 4}[tr({\cal T} V_\mu)]^2\cma\nn\\
{\bar {\cal L}_1}&=&i {1 \over 2} B_{\mu\nu}tr({\cal T} W^{\mu\nu})\cma\nn\\
{\bar {\cal L}_2}&=&i {1  \over 2} B_{\mu\nu}tr({\cal T} [V^\mu,V^\nu])\cma\nn\\
{\bar {\cal L}_3}&=&i           tr( W_{\mu\nu}[V^\mu,V^\nu])\cma\nn\\
{\bar {\cal L}_4}&=&             [tr(V_\mu V_\nu)]^2\cma\nn\\
{\bar {\cal L}_5}&=&             [tr(V_\mu V^\mu)]^2\cma\nn\\
{\bar {\cal L}_6}&=&             tr(V_\mu V_\nu)tr({\cal T} V^\mu)tr({\cal T} V^\nu)\cma\nn\\
{\bar {\cal L}_7}&=&             tr(V_\mu V^\mu)[tr({\cal T} V^\nu)]^2\cma\nn\\
{\bar {\cal L}_8}&=&  {1\over 4} [tr({\cal T} W_{\mu\nu})]^2\cma\nn\\
{\bar {\cal L}_9}&=&i {1  \over 2} tr({\cal T} W_{\mu\nu})tr({\cal T} [V^\mu,V^\nu])\cma\nn\\
{\bar {\cal L}_{10}}&=&             [tr({\cal T} V_\mu)tr({\cal T} V_\nu)]^2\,.
\label{ewcle}
\eea
where the auxiliary variable $V_{\mu}$ and ${\cal T}$ is defined as
\bea
V_{\mu}&=&U^{\dagger}\, D_{\mu}\, U =  U^{\dagger} (\partial_{\mu} - i W_{\mu}^a T^a) U  + i B_{\mu} T^3\,.\label{defd}
\\
{\cal T}&=& 2 U^{\dagger} T^3 U=U^{\dagger} {\tau}^3 U\,,
\eea
with the ${\tau}^3$ is the third Pauli matrices.
For the sake of convenience to refer, we can divide the parameters 
in the EWCL into four groups:
1) $g$, $g'$ and $v$, gauge couplings and vacuum expectation value
respectively; 2) $\al_1$, $\al_8$ and $\beta$, quadratic
anomalous couplings; 3) $\al_2$, $\al_3$, and $\al_9$,
triple anomalous couplings;
4) $\al_4$, $\al_5$, $\al_6$, $\al_7$, and $\al_9$,
quartic anomalous couplings.

Here we would like to point out that due to the different definition 
in the covariant differential
operator $D_{\mu} U$ as given in Eq. (\ref{defd}), 
the signs of triple anomalous couplings
is different than those given in \cite{appel}, while
the quadratic and quartic couplings have the same signs.
The operators ${\bar H_1}$, ${\bar H_2}$, and ${\bar {\cal L}_i},\, i=1,\,\cdots,\,10$
contribute the kinetic, trilinear, and quartic interactions.
While operators ${\bar {\cal L}_{WZ}}$ and ${\bar {\cal L}_0}$
contribute to the mass terms.

Compared with the canonical definition of the
covariant differential operator $D_{\mu}$,
we have absorbed the gauge couplings in the definition of
vector fields. The advantage of this definition is
that it is much easier to define the counter terms. 
While the canonical definition (with
the scaling transformation $W (B) \rightarrow g \, W (g'\,B)$)
complicates the counter term structure.

\indent {\bf Step 2:} By using the background field method \cite{bfm}, we separate both
the vector and Goldstone bosons into the classic and
quantum parts. With the help of Stueckelberg transformation \cite{stuckelberg},
the classic Goldstone can be absorbed by redefining the
classic vector classic field, therefore does not appear in
the computational procedure. At the same time, 
the gauge invariance is guaranteed 
at every computational step, due to the fact that the
Stueckelberg field is an invariant combination with classic vector 
and Goldstone bosons under the background gauge transformation.

The Euler-Lagrange equation of motion of 
the classic fields in mass eigenstates
can be deduced as
\bea
C_1 \partial_{\nu} A^{\mn} + C_2 \partial_{\nu} Z^{\mn}
- e (W^+_{\nu} W^{\mn} - W^-_{\nu} W^{+,\mn})
- {i \over 2} C_5 \partial_{\nu} F_z^{\mn}
\brk + {e \over 2} C_7 (Z\cdot W^+ W^{-,\mu} + Z \cdot W^- W^{+,\mu}
- 2 W^+ \cdot W^- Z^{\mu} ) = 0\,,\\
C_3 \partial_{\nu} Z^{\mn} + C_2  \partial_{\nu} A^{\mn}
- {i \over 2} C_6 \partial_{\nu} F_z^{\mn}
+ {1 \over 2} C_7 (W^+_{\nu} W^{\mn} - W^-_{\nu} W^{+,\mn})
- 2 C_8 W^+ \cdot W^- Z^{\mu} \brk
-   C_9 Z_{\nu} (W^{+,\mu}  W^{-,\nu} + W^{+,\nu}  W^{-,\mu})
- 4 C_{10} Z\cdot Z Z^{\mu}
+ {G^2 \rho v^2 \over 4} Z^{\mu} = 0 \,,\\
d_{\nu} W^{-,\mn} - {i \over 2} C_5 W^-_{\nu} A^{\mn}
- {i \over 2} C_6 W^-_{\nu} Z^{\mn}
- {i \over 2} C_7 d_{\nu} F^{-,\mn}
- {i \over 2} C_7 Z_{\nu} W^{-,\mn} 
- C_8 Z \cdot Z W^{-,\mu} \brk
- C_9 W^- \cdot Z Z^{\mu}
- 2 C_{11} W^{-,\mu} W^{+} \cdot W^{-}
- 2 C_{12} W^{+,\mu} W^{-} \cdot W^{-}
+ {g^2 v^2 \over 4} W^{-,\mu} =0 \,,
\eea
Where $\rho = 1 + 2 \beta$. The equation of motions 
are necessary to eliminate terms
$\partial \cdot Z$ and $d \cdot W^{\pm}$ when we consider
one-loop corrections.

The general renormalized effective 
generating functional can be expressed as
\bea
\Gamma^{eff}({ W},{B},{U}) 
= \Gamma^{tree}({ W},{ B},{ U}) 
+ \de \Gamma({ W},{ B},{ U})
+ \Gamma^{loop}({ W},{ B},{ U})
\eea
Where $\Gamma^{tree}(W,B,U)={\cal L}_{EW}$ 
given in Eqs. (\ref{ewla}---\ref{ewcle}).
Up to one-loop level, the counter term of the EWCL 
$\de \Gamma({W},{B},{ U})$
can be simply formulated as
\bea
\de \Gamma(W,B,U)&=&
2 {\de g \over g} {\bar H_1} + 2 {\de g' \over g'} {\bar H_2}
+ {v \de v \over 2} tr(V\cdot V)
\jsep + { \de \beta v^2 + 2 \beta v \de v \over 4} [tr({\cal T} V_\mu)]^2
+ \sum_{i=1}^{10} \de \alpha_i {\bar {\cal L}^i}
\label{ctt}
\eea

Here we have used one of the advantages of the background field
method that the renormalization constants of classic 
vector field can always be set as $1$. The underlying reason for this
advantage is that there are enough counter terms to absorb
all divergence, the renormalization constant of the classic
vector field is a redundancy. However, when we go beyond
the one-loop level, the renormalization constant for 
vector field is necessary for the one-loop counter terms of 
quantum vector fields.

While $\Gamma^{loop}(W,B,U)$ can be expressed as
\bea
\exp\{ \Gamma^{loop}(W,B,U) \} &=& \int_{{\widehat W} \, {\widehat B} \, \xi \,c}
\exp\{ \int_x {\cal L}({W},{B},{U}; {\widehat W} , {\widehat B} , \xi ,c) \}\,,
\eea
which can be explicitly and systematically calculated to any loop level
by using Feynman-Dayson expansion. However, up to one-loop level, 
the calculation becomes much simplified, since only the quadratic terms in  
${\cal L}({W},{B},{U}; {\widehat W} ,{\widehat B}, \xi ,c)$
contribute. Formally, we can directly perform the Gaussian integral
to get the $\Gamma^{loop}(W,B,U)$.

\indent {\bf Step 3:} In order to perform the path integral in the mass eigenstates,
we rescale the fields (both the classic and quantum fields) 
with their corresponding coupling constant
\bea
W \rightarrow g W\,,\,\, B \rightarrow g' B\,,
\eea
and use the following relation between the mass and interaction
eigenstates
\bea
A & =&  \sin \theta_W W^{3,s} + \cos \theta_W B^{s},\,\,
Z  = -\cos \theta_W W^{3,s} + \sin \theta_W B^{s},\,\,\nnb\\
W^+&=&{1\over \sqrt{2}}(W^{1,s} - i W^{2,s}),\,\,
W^-={1\over \sqrt{2}}(W^{1,s} + i W^{2,s}),\,\,\nnb\\
e  &= &{g' g \over G},\,\,
\tan \theta_{W} ={g' \over g}
\,,
\eea
where the angle $\theta_w$ is the Weinberg angle
and $W^{i,s}$ and $B^{s}$ are Stuckerlberg fields
which have combined both the background vector and
Goldstone fields. 

The rescaling and transformation from the mass and 
interaction eigenstates do not nontrivially change 
the measure of functional
integral. 

We find that to compute in this
way the remnant $U(1)$ gauge invariant is helpful and
useful. Due to this $U(1)$ symmetry, terms 
with $A$ can only exist in $A^{\mn}$ and
$d_{\mu}= \partial - i Q_e e A_{\mu}$. Using this fact,
we can simplify our treatment to $A$ to a certain degree.
We also use this explicit $U(1)$ gauge symmetry to
check our calculation in each step.

However, we can also perform 
the rescaling and transformation
only for the quantum parts, the 
results for both treatments 
should yield the same results.

\indent {\bf Step 4:} After performing the functional Gaussian integral,
we arrive at
\bea
\exp\{ \Gamma^{loop}(W,B,U) \} &=& \exp\{ \Gamma^{loop}(W^{\pm},Z,A) \}\,,\nnb\\
&=&\int \exp \{ - \int_x 
{\cal L}(W^{\pm}, Z, A; {\widehat W^{\pm}} \,, {\widehat Z} \,, {\widehat A}\,,
 \xi^{\pm}, \xi_Z \,, c^{\pm}, c_Z, c_A) \}\,.
\eea
and
\bea
\Gamma^{loop}&=& - {1\over 2}  \left \{ 
   Tr\ln\Box_{V}
+  Tr\ln\Box'_{\xi}
-2 Tr\ln\Box_{c}
\ssep+ Tr\ln\left ( 1 - 
{\stackrel{\rightharpoonup}{X}^{\mu}} \Box^{-1}_{V;\mu\nu} 
{\stackrel{\leftharpoonup}{X}^{\nu}} \Box^{'-1}_{\xi} \right  )
\right \}
\nnb \\
&=& - {1 \over 2} \left \{  
Tr\ln\Box_{V} +   Tr\ln\Box_{\xi} - 2 Tr\ln \Box_{c} 
\ssep + Tr(X^{\alpha \beta} d_\alpha d_\beta \Box_{\xi}^{-1})
- {1 \over 2} Tr(X^{\alpha \beta} d_\alpha d_\beta \Box_{\xi}^{-1} X^{\alpha' \beta'} d_{\alpha'} d_{\beta'} \Box_{\xi}^{-1})
\ssep - Tr( {\stackrel{\rightharpoonup}{X}^{\mu}} 
\Box^{-1}_{V;\mu\nu} 
{\stackrel{\leftharpoonup}{X}^{\nu}} \Box^{-1}_{\xi} )
+ Tr({\stackrel{\rightharpoonup}{X}^{\mu}} 
\Box^{-1}_{V;\mu\nu} 
{\stackrel{\leftharpoonup}{X}^{\nu}} \Box^{-1}_{\xi} 
\, X^{\alpha \beta} d_\alpha d_\beta \Box_{\xi}^{-1} )
\ssep - {1 \over 2} Tr( 
{\stackrel{\rightharpoonup}{X}^{\mu}} \Box^{-1}_{V;\mu\nu} 
{\stackrel{\leftharpoonup}{X}^{\nu}} \Box^{-1}_{\xi} 
{\stackrel{\rightharpoonup}{X}^{\mu'}} \Box^{-1}_{V;\mu'\nu'} 
{\stackrel{\leftharpoonup}{X}^{\nu'}} \Box^{-1}_{\xi})
+ \cdots
\right \} \,\,.
\label{logtrew}
\eea
Here $Tr$ is over all Lorentz and group indices and over all 
coordinate points. Then the contributions of vector bosons, 
Goldstone bosons, and Ghost, have been compactly expressed. 
Each of terms in this expression can be expanded and 
corresponded to a set of gauge independent Feynman diagrams.

\indent {\bf Step 5:} After using the heat kernel and short distance 
expansion technique \cite{hk,sdem}, we can extract all divergences in
the $\Gamma^{loop}(W^{\pm},Z,A)$ and expressed them in the mass eigenstate
basis. The independent complete
operators up to $O(p^4)$ in the mass eigenstate basis
are listed as
\bea
L_{EW}&=&-\sum_{i=1}^{4} C_i O_i+\sum_{i=5}^{12} C_i O_i - O_{M_W} - \rho O_{M_Z} \cma\\
O_1 & = &{1\over 4} A_{\mu\nu} A^{\mu\nu}\cma\nn\\
O_2 & = &{1\over 2} A_{\mu\nu} Z^{\mu\nu}\cma\nn\\
O_3 & = &{1\over 4} Z_{\mu\nu} Z^{\mu\nu}\cma\nn\\
O_4 & = &{1 \over 2} W^{+}_{\mu\nu} W^{-\mu\nu}\cma\nn\\
O_5 & = &{i \over 2} A_{\mu\nu} W^{+\mu} W^{-\nu}\cma\nn\\
O_6 & = &{i \over 2} Z_{\mu\nu} W^{+\mu} W^{-\nu}\cma\nn\\
O_7 & = &{i \over 2} \sbra{W^{+}_{\mu\nu} Z^{\mu} W^{-\nu}
	- W^{-}_{\mu\nu} Z^{\mu} W^{+\nu}}\cma\nn\\
O_8 & = & Z \cdot Z W^{+} \cdot W^{-}\cma\nn\\
O_9 & = & Z \cdot W^{+} Z \cdot W^{-}\cma\nn\\
O_{10} & = & Z \cdot Z Z \cdot Z\cma\nn\\
O_{11} & = & W^{+} \cdot W^{-}  W^{+} \cdot W^{-}\cma\nn\\
O_{12} & = & W^{+} \cdot W^{+}  W^{-} \cdot W^{-}\cma\nn\\
O_{M_W}& = & {v^2 \over 4} W^{+} \cdot W^{-}\cma\nn\\
O_{M_Z} &= & {v^2 \over 8} Z \cdot Z\cma\,.
\eea
This set of operators is explicitly $U(1)$ gauge invariant,
and is the complete operator set up to $O(p^4)$
in mass eigenstates. 

Then the $\Gamma^{loop}(W^{\pm},Z,A)$ is expressed as
the combination of these independent operators, of which
the coefficients are the functions of effective couplings $C_i$:
\bea
\Gamma^{loop}&=&\sum_{i=1} \de C_i (C_i) O^i\,.
\eea

To construct the counter term, we need to restore from
the operators in mass eigenstate basis back to the operators
in electroweak interaction eigenstate basis. Therefore
we use the relation between the operators 
in the mass and interaction eigenstates, which
is given as
\bea
O_{M_Z}&=&-{1\over 2} {\cal L}_0\cma\nn\\
O_{M_W}&=&- {\cal L}_{WZ} + {1 \over 2} {\cal L}_0\cma\nn\\
O_1&=&\wpng{g^4 H_2 + g^2 ({\cal L}_1 - {\cal L}_2) \ssep + g^{'2} (-{\cal L}_4 + {\cal L}_5
+ {\cal L}_6 - {\cal L}_7 + {\cal L}_8 - {\cal L}_9)}{g^2 G^2}\cma\nnb\\
O_2&=&\bpng{g^2 (2 g^{'2} H_2 - {\cal L}_1 + {\cal L}_2) \wsep +
    g^{'2} \left [ {\cal L}_1 - {\cal L}_2 + 2 ({\cal L}_4
- {\cal L}_5 - {\cal L}_6 + {\cal L}_7 - {\cal L}_8 + {\cal L}_9)\right ]}{g G^2 g^{'}}\cma\nnb\\
O_3&=&\wpng{g^{'2} H_2 - {\cal L}_1 + {\cal L}_2 - {\cal L}_4 + {\cal L}_5
+ {\cal L}_6 - {\cal L}_7 + {\cal L}_8 - {\cal L}_9}{G^2}\cma\nnb\\
O_4&=&-\mpng{g^4 ({\cal L}_6 - {\cal L}_7) + G^4 {\cal L}_8
- g^2 G^2 (G^2 H_1 - {\cal L}_3 + {\cal L}_9)}{g^2 G^4}\cma\nnb\\
O_5&=&-\mpng{g^2 {\cal L}_2 + g^{'2} (2 {\cal L}_4 - 2 {\cal L}_5
- 2 {\cal L}_6 + 2 {\cal L}_7 + {\cal L}_9)}{2 g^3 G g^{'}}\cma\nnb\\
O_6&=&\wpng{-{\cal L}_2 + 2 {\cal L}_4 - 2 {\cal L}_5
- 2 {\cal L}_6 + 2 {\cal L}_7 + {\cal L}_9}{2 g^2 G}\cma\nnb\\
O_7&=&-\mpng{2 g^2 ({\cal L}_6 - {\cal L}_7)
+ G^2 ({\cal L}_3 - {\cal L}_9)}{2 g^2 G^3}\cma\nnb\\
O_8&=&\wpng{{\cal L}_7 - {\cal L}_a}{g^2 G^2}\cma\nnb\\
O_9&=&\wpng{{\cal L}_6 - {\cal L}_a}{g^2 G^2}\cma\nnb\\
O_{10}&=&\wpng{2 {\cal L}_a}{G^4}\cma\nnb\\
O_{11}&=&\wpng{2 {\cal L}_5 - 2 {\cal L}_7 + {\cal L}_a}{2 g^4}\cma\nnb\\
O_{12}&=&\wpng{4 {\cal L}_4 - 2 {\cal L}_5 - 4 {\cal L}_6
+ 2 {\cal L}_7 + {\cal L}_a}{2 g^4}\,.
\label{op4}
\eea
Here $H_i$ and ${\cal L}_i$ are operators defined in
in the canonical vector boson fields $W$ and $B$, 
and are related with
${\bar H_i}$ and ${\bar {\cal L}_i}$ defined in
Eqs. (\ref{Lh1}---\ref{ewcle}) by the scaling
transformation of the fields: $g\, W \rightarrow W$ 
and $g^{\prime} B \rightarrow B$.

These relations demonstrate how we can construct the
effective Lagrangian in mass eigenstates and then by
using the inverse Stueckelberg transformation 
to reach EWCL, as prescribed in reference \cite{knetter}.

Using this relation, we obtain $\Gamma^{loop}(W,B,U;C_i)$, 
which is expressed in the independent basis of weak 
interaction eigenstates.
\bea
\Gamma^{loop}&=&\de Z_{H_1} (C_i) {\bar H_1} + \de Z_{H_2} (C_i) {\bar H_2} 
+\de Z_{{\cal L}_{WZ}}(C_i) {\bar {\cal L}_{WZ}}
+\de Z_{{\cal L}_0} {\bar {\cal L}_0} \wsep 
+ \sum_{i=1}^{10} \de Z_{{\cal L}_i} (C_i) {\bar {\cal L}^i}\,.
\eea

And the relation between the effective coupling $C_i$
and the anomalous couplings in the interaction 
eigenstate basis is given as
\bea
\label{c2aa}
 C_1 &=& 1 - \fwpng{g^2 g^{'2}} {2 \alpha_1 + \alpha_8} {G^2}, \nnb \\
 C_2 &=& \fwpng{g g^{'}} {\alpha_1 g^2 - \alpha_1 g^{'2} + \alpha_8 g^2 }{G^2},\nnb \\
 C_3 &=& 1 - \fwpng{g^2}{\alpha_8 g^2 - 2 \alpha_1 g^{'2}} {G^2},\nnb \\
 C_4 &=& 1,\nnb \\
 C_5 &=& \fwpng{2 g g^{'}} {1 - (\alpha_1 + \alpha_2 + \alpha_3 +
    \alpha_8  + \alpha_9 ) g^2 } {G},\nnb \\
 C_6 &=& - \fwpng{2 g^2} { 1 -  (\alpha_3 + \alpha_8 + \alpha_9 )g^2 + (\alpha_1
    + \alpha_2) g^{'2}}{G},\nnb \\
 C_7 &=& \fwpng{2 g^2 }{1 - \alpha_3 G^2}{G},\nnb \\
 C_8 &=& -\frac{g^4}{G^2} + 2 \alpha_3 g^4  + (\alpha_5 + \alpha_7) g^2 G^2,\nnb \\
 C_9 &=&  \frac{g^4}{G^2} - 2 \alpha_3 g^4  + (\alpha_4 + \alpha_6) g^2 G^2,\nnb \\
 C_{10} &=& \fwpng{G^4} {\alpha_4 + \alpha_5 + 2 \alpha_6 + 2 \alpha_7 + 2 \alpha_a}{4},\nnb \\
 C_{11} &=& -\frac{g^2}{2} + \frac{g^4}{2} \left ( 2 \alpha_3 + \alpha_4  + 2 \alpha_5  + \alpha_8 + 2 \alpha_9 \right ),\nnb \\
 C_{12} &=&  \frac{g^2}{2} - \frac{g^4}{2} \left ( 2 \alpha_3 - \alpha_4  + \alpha_8  + 2 \alpha_9  \right )\,.
\label{c2ab}
\eea
Using this relation, the $\Gamma^{loop}(W,B,U;C_i)$ is 
transformed to be $\Gamma^{loop}(W,B,U;\alpha_i)$:
\bea
\Gamma^{loop}&=&\de Z_{H_1} (\al_i) {\bar H_1} + \de Z_{H_2} (\al_i) {\bar H_2} 
+\de Z_{{\cal L}_{WZ}}(\al_i) {\bar {\cal L}_{WZ}}
+\de Z_{{\cal L}_0}(\al_i) {\bar {\cal L}_0} \wsep 
+ \sum_{i=1}^{10} \de Z_{{\cal L}_i} (\al_i) {\bar {\cal L}^i}\,.
\label{gct}
\eea

\indent {\bf Step 6:} Using the $\Gamma^{loop}(W,B,U;\alpha_i)$
given in Eq. (\ref{gct}) and the counter ctructure given in Eq. (\ref{ctt}), 
it is straightforward
to construct counter terms $\de \Gamma^{loop}(W,B,U;\alpha_i)$.
With the constructed counter term, we arrive at the 
RGEs, which at one-loop level can be
expressed as the general form:
\bea
{d \, c \over d t} &=& {1 \over 8 \pi^2 } \beta_c\,.
\eea
where the complete $\beta_c$ functions of
each coupling of the EWCL are quite complicated ( which
will be provided in our full paper \cite{longppr}).

In order to make the RGEs ease to use, 
we expand the effective couplings $C_i$
and keep only the linear terms of anomalous couplings $\alpha_i$
in the $\beta$ functions,
then we get the following simplified version of $\beta$ 
functions of couplings in the EWCL
\bea
\label{rge0}
\beta_{g}&=&
g^2 \bbbkf{ -\nkf{ \frac{29}{4} }  - \frac{\beta}{6} - 
    \my{\al_1} {\my{g'}}^2 - 
    4 \my{\al_8} g^2   + 
    \frac{5 \my{\al_2} {\my{g'}}^2}{6} + 
    \my{\al_3} \sbbkf{ \frac{50 g^2}{3} - 
       \frac{3 {\my{g'}}^2}{2} }  
   + \frac{23 \my{\al_9} g^2}{6}} 
\label{betag}\,,\\
\beta_{g'}&=&
  {\my{g'}}^2
\sbbkf{ \frac{1}{12} - \frac{\beta}{3} - 2 \my{\al_1} g^2 - 
    3 \my{\al_2} g^2 + \frac{5 \my{\al_3} g^2}{3} }  
\label{betagp}\,,\\
\beta_{\al_1}&=&
\frac{1}{12} + 4 \my{\al_1} g^2 - \my{\al_8} g^2
  + \frac{9 \my{\al_2} g^2}{2} - 
  \frac{37 \my{\al_3} g^2}{6}  - 
  \frac{3 \my{\al_9} g^2}{2}
\,,\\
\beta_{\al_2}&=&
-\nkf{ \frac{1}{24} }  - \frac{\beta}{6} - 
  \frac{5 \my{\al_1} g^2}{2} \wsep + 
  \my{\al_2} \sbbkf{ \frac{-{\my{e}}^2}{6} - 2 g^2 + 
     \frac{5 {\my{g'}}^2}{4} } + 
  \my{\al_3} \sbbkf{ \frac{{\my{e}}^2}{3} - 
     \frac{29 g^2}{12} }  + 
  \my{\al_9} \sbbkf{ \frac{{\my{e}}^2}{6} - 
     \frac{25 g^2}{12} }  \wsep + 
  \my{\al_4} \sbbkf{ \frac{3 g^2}{4} + 
     \frac{{\my{g'}}^2}{2} }  + 
  \my{\al_5} \sbbkf{ \frac{-3 g^2}{2} + {\my{g'}}^2 }  + 
  \frac{\my{\al_6} {\my{g'}}^2}{2} + 
  \my{\al_7} {\my{g'}}^2 
\,,\\
\beta_{\al_3}&=&
-\nkf{ \frac{1}{24} }  + \frac{\beta}{6} + 
  \my{\al_1} \sbbkf{ {\my{e}}^2 - 
     \frac{{\my{g'}}^2}{4} }   + 
  \my{\al_8} \sbbkf{ -{\my{e}}^2 + 
     \frac{5 g^2}{4} } \wsep + 
  \my{\al_2} \sbbkf{ \frac{{\my{e}}^2}{6} - 
     \frac{3 {\my{g'}}^2}{4} }  + 
  \my{\al_3} \sbbkf{ \frac{35 {\my{e}}^2}{12} + 
     \frac{61 g^2}{12} }    + 
  \my{\al_9} \sbbkf{ \frac{-{\my{e}}^2}{6} - \frac{g^2}{6} }  \wsep + 
  \my{\al_4} \sbbkf{ \frac{-9 g^2}{4} + 
     \frac{5 {\my{g'}}^2}{8} }+ 
  \my{\al_5} \sbbkf{ \frac{9 g^2}{2} - 
     \frac{{\my{g'}}^2}{4} }    + 
  \my{\al_6} \sbbkf{ \frac{-9 g^2}{4} + 
     \frac{5 {\my{g'}}^2}{8} }+ 
  \my{\al_7} \sbbkf{ \frac{9 g^2}{2} - 
     \frac{{\my{g'}}^2}{4} }   
\,,\\
\beta_{\al_4}&=&
-\nkf{ \frac{1}{12} }  + 
  \my{\al_2} \sbbkf{ \frac{{\my{e}}^2}{3} - 
     \frac{{\my{g'}}^2}{6} }  + 
  \my{\al_3} \sbbkf{ \frac{-2 {\my{e}}^2}{3} + 
     \frac{43 g^2}{6} }  + 
  \my{\al_9} \sbbkf{ \frac{-{\my{e}}^2}{3} + 
     \frac{17 g^2}{2} } \wsep + 
  \my{\al_4} \sbbkf{ \frac{7 g^2}{2} + 4 {\my{g'}}^2 } + 
  \my{\al_5} \sbbkf{ 6 g^2 - {\my{g'}}^2 }  + 
  \my{\al_6} \sbbkf{ \frac{7 g^2}{2} - 
     \frac{3 {\my{g'}}^2}{2} }   - 2 \my{\al_7} {\my{g'}}^2  
\,,\\
\beta_{\al_5}&=&
-\nkf{ \frac{1}{24} }  + \frac{\beta}{2} + 
  \my{\al_2} \sbbkf{ \frac{-{\my{e}}^2}{3} + 
     \frac{2 {\my{g'}}^2}{3} }   + 
  \my{\al_3} \sbbkf{ \frac{2 {\my{e}}^2}{3} - 
     \frac{37 g^2}{6} }  + 
  \my{\al_9} \sbbkf{ \frac{{\my{e}}^2}{3} - 8 g^2 } \wsep + 
  \my{\al_4} \sbbkf{ \frac{3 g^2}{2} - 
     \frac{3 {\my{g'}}^2}{4} }  + 
  \my{\al_5} \sbbkf{ -g^2 + \frac{7 {\my{g'}}^2}{2} } + 
  \my{\al_6} \sbbkf{ \frac{-g^2}{4} + 
     \frac{5 {\my{g'}}^2}{4} }  + 
  \my{\al_7} \sbbkf{ \frac{7 g^2}{2} + 
     \frac{3 {\my{g'}}^2}{2} }  
\,,\\
\beta_{\al_6}&=&
  \my{\al_2} \sbbkf{ \frac{-7 {\my{e}}^2}{6} + 
     \frac{{\my{g'}}^2}{2} - 
     \frac{3 {\my{e}}^2 {\my{g'}}^2}{2 G^2} }  + \my{\al_3} 
   \sbbkf{ -9 {\my{e}}^2 + 
     \frac{3 {\my{e}}^2 {\my{g'}}^2}{G^2} } + 
  \my{\al_9} \sbbkf{ \frac{-4 {\my{e}}^2}{3} - 
     \frac{55 g^2}{6} + \frac{3 {\my{e}}^2 {\my{g'}}^2}
      {2 G^2} } \wsep +
\my{\al_4} \sbbkf{ \frac{3 {\my{e}}^2}{2} - 
     \frac{7 {\my{g'}}^2}{4} }  + 
  \my{\al_5} \sbbkf{ -7 {\my{e}}^2 + 
     \frac{7 {\my{g'}}^2}{2} }   + 
  \my{\al_6} \sbbkf{ \frac{3 {\my{e}}^2}{2} - 
     \frac{7 g^2}{4} + \frac{13 {\my{g'}}^2}{4} } \wsep   + 
  \my{\al_7} \sbbkf{ -7 {\my{e}}^2 + 6 g^2 + 
     4 {\my{g'}}^2 }  + 
  \my{\al_{10}} \sbbkf{ 7 g^2 - {\my{g'}}^2 } 
\,,\\
\beta_{\al_7}&=&
\frac{-3 \beta}{4} + \my{\al_2} 
   \sbbkf{ \frac{7 {\my{e}}^2}{6} - 
     \frac{3 {\my{g'}}^2}{4} + 
     \frac{2 {\my{e}}^2 {\my{g'}}^2}{G^2} } + 
  \my{\al_3} \sbbkf{ 9 {\my{e}}^2 - 
     \frac{4 {\my{e}}^2 {\my{g'}}^2}{G^2} }  + 
  \my{\al_9} \sbbkf{ \frac{4 {\my{e}}^2}{3} + 
     \frac{107 g^2}{12} - \frac{2 {\my{e}}^2 {\my{g'}}^2}
      {G^2} }  \wsep + 
  \my{\al_4} \sbbkf{ \frac{-9 {\my{e}}^2}{2} + 
     \frac{g^2}{2} + \frac{17 {\my{g'}}^2}{8} } + 
  \my{\al_5} \sbbkf{ 4 {\my{e}}^2 - 
     \frac{7 {\my{g'}}^2}{4} }  + 
  \my{\al_6} \sbbkf{ \frac{-9 {\my{e}}^2}{2} + 
     \frac{19 g^2}{8} + \frac{{\my{g'}}^2}{8} } \wsep  + 
  \my{\al_7} \sbbkf{ 4 {\my{e}}^2 - \frac{25 g^2}{4} + 
     \frac{{\my{g'}}^2}{4} }   
+ 3 \my{\al_{10}} g^2 
\,,\\
\beta_{\al_8}&=&
\frac{\beta}{2} + \my{\al_1} {\my{g'}}^2
 + 12 \my{\al_8} g^2  - 
  \frac{5 \my{\al_2} {\my{g'}}^2}{6} + 
  \frac{3 \my{\al_3} {\my{g'}}^2}{2}+ 
  \frac{31 \my{\al_9} g^2}{6} 
\,,\\
\beta_{\al_9}&=&
\frac{-\beta}{2} + 
  \my{\al_1} \sbbkf{ -{\my{e}}^2 + 
     \frac{{\my{g'}}^2}{4} } + 
  \my{\al_8} 
   \sbbkf{ {\my{e}}^2 - \frac{15 g^2}{4} }  \wsep + 
  \frac{\my{\al_2} {\my{g'}}^2}{3}  + 
  \my{\al_3} \sbbkf{ \frac{-13 {\my{e}}^2}{4} - 
     \frac{{\my{g'}}^2}{2} }   + 
  \my{\al_9} \sbbkf{ \frac{67 g^2}{12} + 
     \frac{3 {\my{g'}}^2}{2} } \wsep - 
  \frac{9 \my{\al_4} {\my{g'}}^2}{8} - 
  \frac{3 \my{\al_5} {\my{g'}}^2}{4} + 
  \my{\al_6} \sbbkf{ \frac{9 g^2}{4} - 
     \frac{9 {\my{g'}}^2}{8} }  + 
  \my{\al_7} \sbbkf{ \frac{-9 g^2}{2} - 
     \frac{3 {\my{g'}}^2}{4} } 
\,,\\
\beta_{\al_{10}}&=&
- \frac{\nkf{ \my{\al_2} {\my{e}}^2 {\my{g'}}^2 } }{2 G^2} 
+ \frac{\my{\al_3} {\my{e}}^2 
     {\my{g'}}^2}{G^2} + 
  \frac{\my{\al_9} {\my{e}}^2 {\my{g'}}^2}{2 G^2} \wsep + 
  \my{\al_4} \sbbkf{ 3 {\my{e}}^2 - 
     \frac{3 g^2}{4} - 2 {\my{g'}}^2 }  + 
  \my{\al_5} \sbbkf{ 3 {\my{e}}^2 - 3 {\my{g'}}^2 } + 
  \my{\al_6} \sbbkf{ 3 {\my{e}}^2 - g^2 - 
     \frac{13 {\my{g'}}^2}{4} } \wsep + 
  \my{\al_7} \sbbkf{ 3 {\my{e}}^2 - 4 {\my{g'}}^2 }    + 
  \my{\al_{10}} \sbbkf{ -12 g^2 + {\my{g'}}^2 } 
\,,\\
\beta_{v}&=&
\frac{3 g^2}{4} + \frac{3 {\my{g'}}^2}{8} + 
  \frac{5 \beta g^2}{4} - 
  \frac{11 \my{\al_1} g^2 {\my{g'}}^2}{4} + 
  \frac{3 \my{\al_8} g^4}{8} \wsep  + 
  \my{\al_2} \sbbkf{ \frac{-5 g^2 {\my{g'}}^2}{2} + 
     \frac{3 {\my{g'}}^4}{4} } + 
  \my{\al_3} \sbbkf{ g^4 - \frac{5 g^2 {\my{g'}}^2}{2} }  +
  \my{\al_9} \sbbkf{ \frac{g^4}{2} - 
     \frac{3 g^2 {\my{g'}}^2}{4} }  \wsep + 
  \my{\al_4} \sbbkf{ \frac{-7 g^4}{2} - 
     \frac{3 g^2 {\my{g'}}^2}{4} - \frac{3 {\my{g'}}^4}{8} }+ 
  \my{\al_5} \sbbkf{ \frac{-27 g^4}{4} - 4 g^2 {\my{g'}}^2 - 
     2 {\my{g'}}^4 } \wsep  - \my{\al_6} {8 G^4 \over 3} -2 
  \my{\al_7}  G^4    
\,,\\
\beta_{\beta}&=&
\frac{-15 \beta g^2}{4} - 
  \frac{3 {\my{g'}}^2}{8} - \frac{3 \beta {\my{g'}}^2}{4} + 
  \frac{11 \my{\al_1} g^2 {\my{g'}}^2}{4}  -
  \frac{3 \my{\al_8} g^4}{8} \wsep + 
  \my{\al_2} \sbbkf{ \frac{5 g^2 {\my{g'}}^2}{2} - 
     \frac{3 {\my{g'}}^4}{4} } + 
  \frac{5 \my{\al_3} g^2 {\my{g'}}^2}{2} + 
  \my{\al_9} \sbbkf{ \frac{-g^4}{2} + 
     \frac{3 g^2 {\my{g'}}^2}{4} }  \wsep    + 
  \my{\al_4} \sbbkf{ \frac{-19 g^2 {\my{g'}}^2}{4} - 
     \frac{19 {\my{g'}}^4}{8} }  + 
  \my{\al_5} \sbbkf{ \frac{-3 g^2 {\my{g'}}^2}{2} - 
     \frac{3 {\my{g'}}^4}{4} }  + \my{\al_6} \sbbkf{ \frac{-47 g^4}{8} - 
     \frac{41 g^2 {\my{g'}}^2}{4} - 
     \frac{41 {\my{g'}}^4}{8} } \wsep + 
  \my{\al_7} \sbbkf{ \frac{-15 g^4}{2} - 7 g^2 {\my{g'}}^2 - 
     \frac{7 {\my{g'}}^4}{2} } - 
{11 \over 2}  \my{\al_{10}} G^4
\,.
\label{rge1}
\eea

There are several comments on the $\beta$ functions in order:
1) The $\al_i$ always appear with $g^2$, $g'$, $e^2$, and $G^2$.
If we count the couplings of operators with mass dimension 4 as
$-2$, then the combination of $\al_i g^2$ is $O(1)$.
2) Although $\al_1$, $\al_8$ and $\beta$ belong to 
the quadratic anomalous, the $\beta$ function of $\beta$ 
parameter receives
the radiative corrections from quartic anomalous couplings
while those of $\al_1$ and $\al_8$ do not. The reason is
easy to understand in Feynman diagram as given in Figure 1.
The $\al_1$ and $\al_8$
can only receive radiative corrections through the first
diagram, Fig. 1(a), while $\beta$ can 
receive radiative corrections from both diagrams, Fig. 1(a) and 1(b).
It is Fig. 1(b) that makes the quartic couplings contribute directly
to $\beta$ parameter. 3) Due to its large numerical factor, $\al_3$ can
affect much more than the rest to the running of $g$.
Similarly, due to the large numerical factor 
in $\beta_{\al_1}$, $\al_2$ and $\al_3$ can 
affect the running of $\al_1$ significantly.
4) If we use the naive dimensional analysis from \cite{manahar} and
assume that all anomalous couplings are of the one-loop
corrections of $O(p^2)$ operators, then all terms of $\al_i$ and
$\beta$ in $\beta$ function should belong to two-loop 
order. At the one-loop
level, we can neglect them, then we
reach to the previous results obtained in \cite{long,bernard}:
\bea
\beta_{g}&=& {g^3 \over 2} \sbbkf{ - \frac{29}{4} } \,,\nnb\\
\beta_{g'}&=& {{\my{g'}}^3 \over 2} \sbbkf{ \nkf{ \frac{1}{12} }  
  }  \,,\nnb\\
\beta_{\al_1}&=& \nkf{ \frac{1}{12} } \,,\nnb\\
\beta_{\al_2}&=&-\frac{1}{24} \,,\nnb\\
\beta_{\al_3}&=& -\frac{1}{24} \,,\nnb\\
\beta_{\al_4}&=&-\frac{1}{12}   \,,\nnb\\
\beta_{\al_5}&=& -\frac{1}{24}  \,,\nnb\\
\beta_{\al_6}&=&0\,,\nnb\\
\beta_{\al_7}&=&0\,,\nnb\\
\beta_{\al_8}&=&0\,,\nnb\\
\beta_{\al_9}&=& 0\,,\nnb\\
\beta_{\al_{10}}&=& 0\,,\nnb\\
\beta_{v}&=&v \bbbkf{ \frac{3 g^2}{4} + \frac{3 {\my{g'}}^2}{8} } \,,\nnb\\
\beta_{\beta}&=&-\frac{3 {\my{g'}}^2}{8} 
\,.
\eea
To reach this previous result is also one of the  
foolproof checks for our calculation procedure. This
result comes from the contribution of pure Goldstone
boson loops. 
5) Compared with the simplified version of RGEs we have obtained
previously in \cite{ewclrge}, this version is even simpler.
6) It might be helpful to express the simplified version of RGEs
into the linear matrix form (higher order nonlinear terms have been
dropped):
\bea
{d  \over d t} \{ \al_i \}= {1 \over 8 \pi^2} \mbkf{ \{ C_{\al_i} \} + \beta_\gamma \{ \al_i \}}
\,,
\eea
where the $\{ \al_i \}^T$ is defined as 
$\{\al_1,...,\al_{10}, \beta\}$, 
and the constant $\{ C_{\al_i} \}^T=\{-1/12, 1/24, 
1/24, 1/12, 1/24, 0, 0, 0, 0, 0, 3 {\my{g'}}^2/8 \}$.
With the expression given in Eqs.(\ref{rge0}---\ref{rge1}), 
it is straightforward to determine the matrix $\beta_\gamma$,
which just indicates mixings between the anomalous operators.

Below we study some phenomenological applications of the RGEs.
We take $m_Z(m_Z)=91.187$ GeV,
$g_(m_Z)=0.651$, $g'(m_Z)=0.357$, and $v(m_Z)=246.708$ GeV
as some of inputs.

We consider the constraints of the precision test parameters
to the anomalous couplings ($\al_1$, $\al_8$, and $\beta$) 
at the cutoff energy scale.
The relations \cite{appel} between the quadratic anomalous
couplings and the oblique precision test parameter 
$S$, $T$ and $U$ at $\mu=m_Z$ are given as
 \bea
S&=& - 16 \pi \alpha_1\,,\\
T&=&{\rho - 1 \over \alpha_{em}}= {2 \beta \over \alpha_{em}}\,,\\
U&=& - 16 \alpha_8\,,
\eea
According to the extant experiment results
\cite{smewfit,bagger,lepexp}, we take the current
constraints on the precision test parameters as
\bea
S&=&-0.13 \pm 0.10\,,\\
T&=&-0.13 \pm 0.11\,,\\
U&=& 0.22 \pm 0.13\,.
\eea
Therefore the lower energy boundary $\al_1(m_Z)$, $\al_8(m_Z)$,
and $\beta(m_Z)$, is determined as our inputs.
However, we would like emphasize that
our theoretical framework assumes no Higgs. 
So these values of  
precision test parameters, which are obtained by fitting
in the standard model with a Higgs, should not be
regarded with too much seriousness.
However, what we want to show below is 
the running behavior of $\al_1$, $\al_8$, and $\beta$
at the ultraviolet cutoff $\Lambda$.

While for other anomalous couplings
we take the following variation range as inputs
\bea
\label{expcon}
(\al_2(m_Z)\,\,\al_3(m_Z)\,\,\al_9(m_Z))\sim O(0.1),\,\,\nnb\\
(\al_4(m_Z)\,\,\al_5(m_Z)\,\,\al_6(m_Z)\,\,\al_7(m_Z)\,\,\al_a(m_Z))\sim O(1)\,.
\eea
These are consistent with the current LEP measurement \cite{smewfit,bagger,lepexp,lep}.

So with these low energy inputs at $\mu=m_Z$ as inputs,
by solving the RGEs, we can study $\al_i(\Lambda)$ with the RGEs, 
which reflect the experimental constraints 
on the unknown possible underlying theories. 

Before the actual numerical analysis, formally, 
we can express the solutions of $\al_i$ from the RGEs
as
\bea
\al_{i}(m_Z)=\al_i(\Lambda) + {1 \over 16 \pi^2} \beta_{\al_i} \ln({\Lambda \over m_Z})^2\,.
\eea
From this formal solution, we know that the low energy
value of $\al_i(m_Z)$ is related with three factors:
its initial value at the matching conditions $\al_i(\Lambda)$,
which is determined by the underlying theory;
$\beta_{\al_i}$ functions, which depends on
the other anomalous couplings and reflects the contributions
of the vector and Goldstone quantum;
the ultraviolet cutoff $\Lambda$, which is 
energy scale where new physics should show up
but is unknown to us. 
We would like to point out two obvious facts about this formal
solutions:  1) Due to the fact that $\beta_{\al_i}$ is
suppressed by loop factor $1/(8 \pi^2)$, if $\al_{i}(m_Z)$
is small (say of order one-loop, like $\al_1$ and $\al_8$), 
then the radiative corrections are relatively important.
2) Generally speaking, for a specific set of $\al_i(m_Z)$, 
the larger the $\Lambda$, the 
larger the variation of $\al_i(\Lambda)$.

We consider three cases with different ultraviolet cutoffs:
case 1, the ultraviolet cutoff is 
set as $\Lambda_{UV}=600$ GeV;
case 2, $\Lambda_{UV}=1000$ GeV;
and case 3, $\Lambda_{UV}=3000$ GeV.

Figures (1---3) are devoted for the first group of 
parameters in EWCL, {\it i.e.} gauge couplings 
($g$ and $g'$)and vacuum expectation value ($v$).
In the Figs. 1(a---c), the correlations between
$\al_3$ and $g$ is shown. Due to the large factor
before $\al_3$ in the $\beta$ function of $g$, 
the $\al_3$ can affect the running of $g$. 
The phenomenological meaning
of this fact is that it might provide an 
alternative way to determine or constrain
the anomalous coupling $\al_3$
if the running of $g$ can be reliably measured
in the future experiments to a precision $10^{-2}$.

While Figs. 2(a---c) show that the running of
$g'$ is quite small and has a very weak correlation
to $\al_3$. Such a difference between $g$ and $g'$
can be traced back to the
$\beta$ functions of $g$ and $g'$ given in 
Eqs. (\ref{betag}---\ref{betagp}).
The direct physics reason includes that the degree of freedoms
of the $U(1)$ representation in EWCL is much fewer than
that of the $SU(2)$ one and the coupling
of $U(1)$ is only half of that $SU(2)$.

Figs. 3(a---c) is devoted to reveal the correlation
between the $\al_5$ and $v$, and due to the
large numerical factor before $\al_5$ in the $\beta$ function
of $v$, the running of $v$ can be affected by the 
value of $\al_5$. As the $\al_3$ in the case $g$, if the running
of $v$ can be measured to a precision of $10^{-2}$,
this will help to constrain quartic couplings, like $\al_5$.

From Figs. 1(d), 2(d) and 3(d), we know that the deviation of
predictions of the simplified RGEs and the complete ones
is small under the scanned region specified in Eq. (\ref{expcon}).

Figures (4---6) are devoted to the quadratic group.
Fig. 4(a---c) shows the correlation of $\al_1(m_Z)$ with
its value at the UV cutoff. Due to the radiative effect
from other anomalous couplings, $\al_1(\Lambda)$ can be
either positive or negative. The direct reason
is the contribution of $\al_2$ and $\al_3$ terms in
the $\beta_{\al_1}$, which have relative large numerical
factors. With the increase of $\Lambda$ the variation range of 
$\al_1(\Lambda)$ increases. 
In \cite{bagger}, the mechanism for the change
of sign $S$ is solely due to the ${1 \over 12}$ in the
$\beta_{\al_1}$, the pure Goldstone contribution; while
here we find the reason for the change
of the sign is due to the combination of this ${1 \over 12}$ 
and the interference of $\al_2$ and $\al_3$. If the interference
is constructive, $\al_1(\Lambda)$ can have a quite large
value with changed sign; if the interference is destructive,
$\al_1(\Lambda)$ can keep its sign.

Fig. 5(a---c) shows the correlation of 
$\al_8(m_Z)$ with its value at the UV cutoff.
Similar to the $\al_1$ case, due to the large contribution
from triple couplings, $\al_8(\Lambda)$ can be either
positive or negative. 

The deviation range of $\al_1(\Lambda)$
and $\al_8(\Lambda)$ from the variation range of
$\al_1(m_Z)$ and $\al_8(m_Z)$ can reach to one order.
Fig. 6(a---c) shows the correlation of $\beta(m_Z)$ with
its value at the UV cutoff. The value of $\beta(\Lambda)$
can deviate from $\beta(m_Z)$ significantly, and
the deviation range from the variation range can reach 
to two orders.

Why is there such a big difference between $\beta$ and
$\al_1$ (and $\al_8$)? The basic reason is that
the $\beta$ can get the direct corrections from the
unknown quartic couplings, which might reach $O(1)$;
while $\al_1$ ($\al_8$) can only get the direct corrections
from triple anomalous couplings. The correction 
of quartic couplings to $\al_1$ ($\al_8$) is indirect via
triple anomalous couplings. This explains the difference.
One interesting phenomenological consequence about the
contributions of direct quartic anomalous couplings 
to $\beta$ parameter is that, if we can measure the running 
of $\beta$ to a certain precision,
this will serve as an good way to constrain the
magnitude of quartic anomalous couplings.

Figs. 4(d) and 5(d) show that the simplified
version of RGEs is as good as the complete one. While for Fig.
6(d), when the magnitude of
$\beta(\Lambda)$ is large ($|\beta(\Lambda)| > 0.1$), the deviation
of the results of the complete RGEs and simplified RGEs becomes large.

In summary, we have shown the main steps and methods of the
computational procedure for the renormalization 
of the EWCL. We arrived at the RGEs. And for the sake of easiness to use, we
provide a simplified version, which, as has been demonstrated,
is quite reliable for the parameter space we have considered.
By using the one loop RGEs of EWCL, we have studied some region of
the permitted parameter space of the EWCL at the ultraviolet cutoff 
by incorporating the current precision test constraints. 
We have found that due to the
radiative corrections from triple anomalous couplings,
$\al_1(\Lambda)$ and $\al_8(\Lambda)$ can have a considerable 
deviation from $\al_1(m_Z)$ and $\al_8(m_Z)$ (which can reach
one order), and can be either
positive or negative. While due to the contributions of
quartic anomalous couplings, $\beta(\Lambda)$ can 
have quite a large deviation from the $\beta(m_Z)$ 
(which can reach two order).

\acknowledgements
The author would like to thank Dr. S. Dutta and Prof. K. Hagiwara
for helpful discussions. The author is aslo indebted to
Dr. H.J. He, Prof. Y. P. Kuang and Prof. Q. Wang,
for some constructive suggestions. The work is 
supported in part the Chinese Postdoctoral Science Foundation
and the CAS K. C. Wong Postdoctoral Research Award Foundation,
and in part by Grant-in-Aid Scientific Research from 
Ministry of Education, Culture, Science and Technology of Japan,
and partially supported by the JSPS fellowship program.

\newpage
{\Large \bf Figures and Captions:}\\
\vskip 2cm 
\begin{figure}
\begin{minipage}[t]{6.8cm}
     \epsfig{file=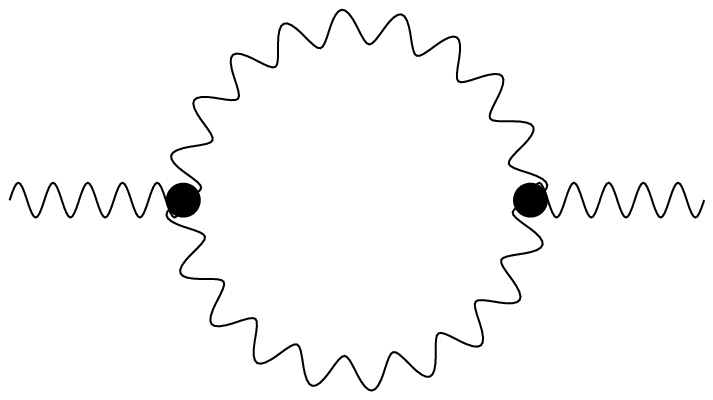,width=6.8cm}
     \mbox{ }\hfill\hspace{1cm}(a)\hfill\mbox{ }
     \end{minipage}
     \hspace{0.2cm}
     \begin{minipage}[t]{6.8cm}
     \epsfig{file=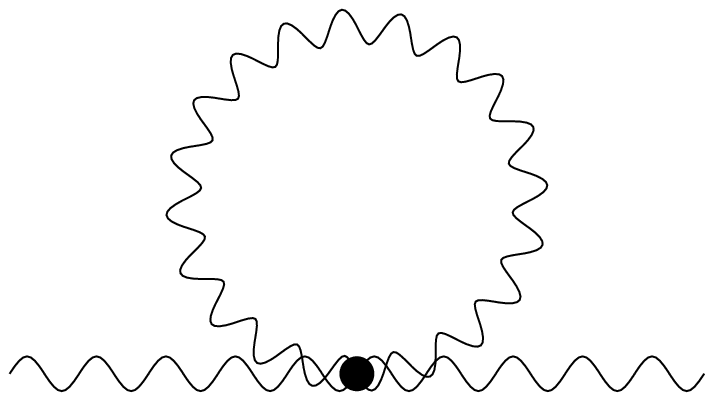,width=6.8cm}
     \mbox{ }\hfill\hspace{1cm}(b)\hfill\mbox{ }
     \end{minipage}
\caption{Two types of Feynman diagrams for the radiative corrections
to quadratic anomalous couplings}
\label{fig0}
\end{figure}

\begin{figure}
\epsfig{file=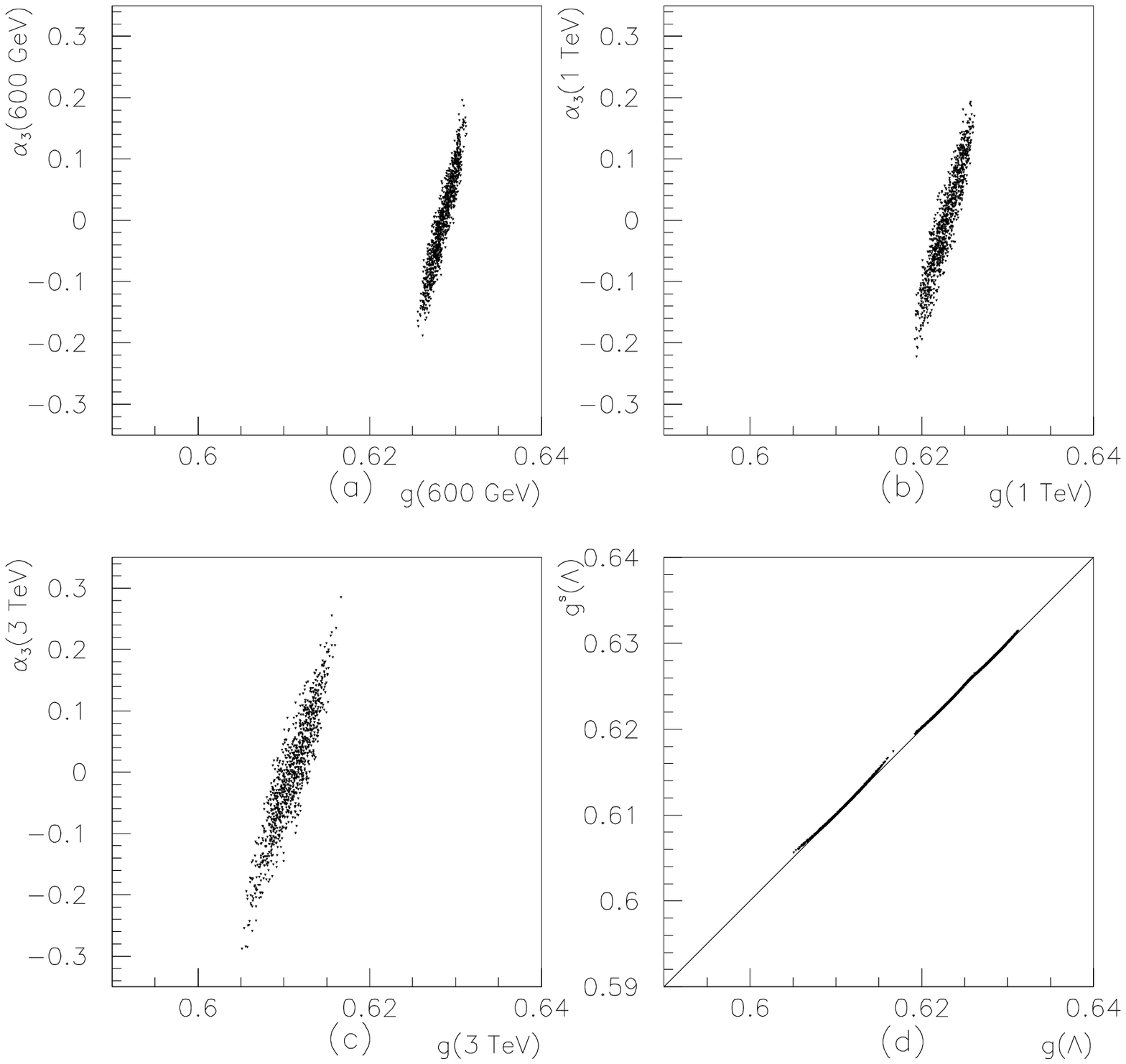,width=\textwidth}
\caption{The correlation between the running of gauge coupling constant $g$
and the anomalous coupling $\al_3$, 
(a) is devoted to case (1) where $\Lambda=600$ GeV; 
(b) is devoted to case (2) where $\Lambda=1 $ TeV; 
(c) is for case (3) where $\Lambda=3$ TeV.
For these three figures, 
x axises are the value of $g(600 GeV)$, $g(1 TeV)$,
and $g(3 TeV)$, respectively; y axises are the
value of $\al_3(600 GeV)$, $\al_3(1 TeV)$,
and $\al_3(3 TeV)$, respectively.
Fig. (d) is devoted to compare the value of $g(\Lambda)$ and
$g^s(\Lambda)$ by solving the complete version and simplified 
version of RGEs, respectively; the line $y=x$ is depicted for the sake
of contrast.}
\label{fig1}
\end{figure}

\begin{figure}
\epsfig{file=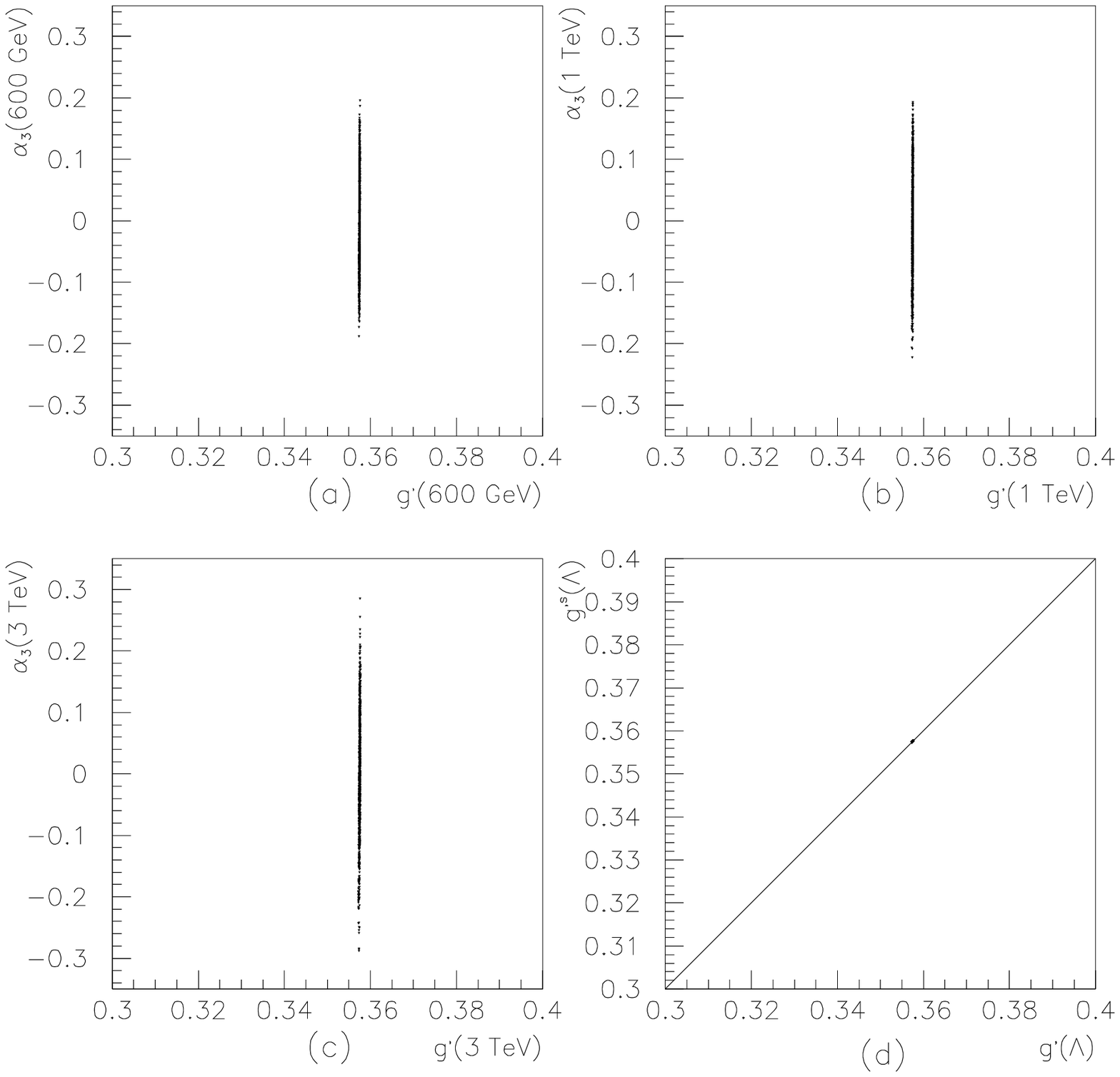,width=\textwidth}
\caption{The correlation between the running of gauge coupling constant $g'$
and the anomalous coupling $\al_3$, 
(a) is devoted to case (1) where $\Lambda=600$ GeV; 
(b) is devoted to case (2) where $\Lambda=1 $ TeV; 
(c) is for case (3) where $\Lambda=3$ TeV.
For these three figures, 
x axises are the value of $g'(600 GeV)$, $g'(1 TeV)$,
and $g'(3 TeV)$, respectively; y axises are the
value of $\al_3(600 GeV)$, $\al_3(1 TeV)$,
and $\al_3(3 TeV)$, respectively.
Fig. (d) is devoted to compare the value of $g^{\prime}(\Lambda)$ and
$g^{\prime s}(\Lambda)$ by solving the complete version and simplified 
version of RGEs, respectively; the line $y=x$ is depicted for the sake
of contrast.}
\label{fig2}
\end{figure}

\begin{figure}
\epsfig{file=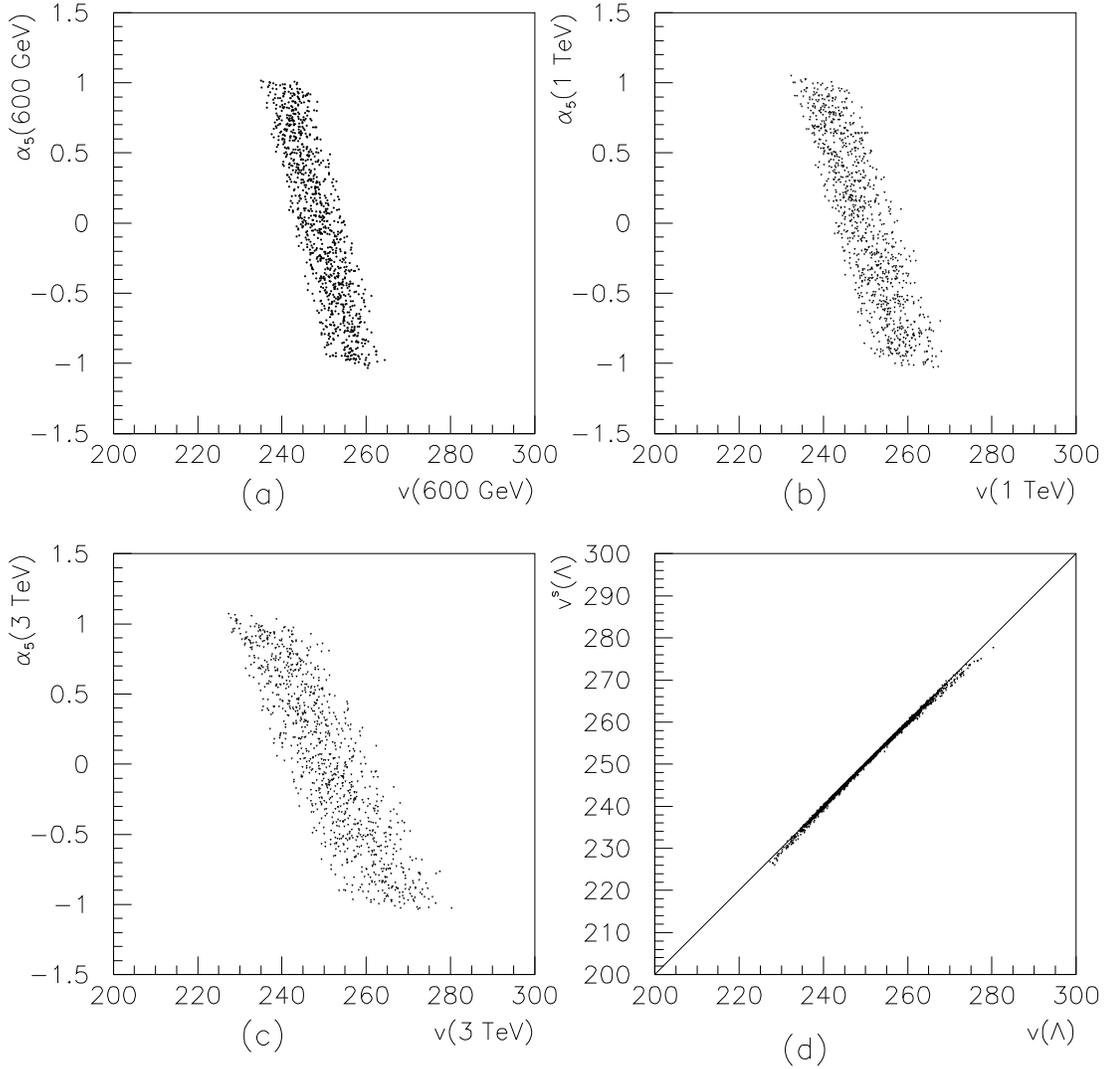,width=\textwidth}
\caption{The correlation between the running 
of vacuum expectation
value $v$ and the anomalous coupling $\al_5$, 
(a) is devoted to case (1) where $\Lambda=600$ GeV; 
(b) is devoted to case (2) where $\Lambda=1 $ TeV; 
(c) is for case (3) where $\Lambda=3$ TeV.
For these three figures, 
x axises are the value of $v(600 GeV)$, $v(1 TeV)$,
and $v(3 TeV)$, respectively; y axises are the
value of $\al_5(600 GeV)$, $\al_5(1 TeV)$,
and $\al_5(3 TeV)$, respectively.
Fig. (d) is devoted to compare the value of $v(\Lambda)$ and
$v^{s}(\Lambda)$ by solving the complete version and simplified 
version of RGEs, respectively; the line $y=x$ is depicted for the sake
of contrast.}
\label{fig3}
\end{figure}

\begin{figure}
\epsfig{file=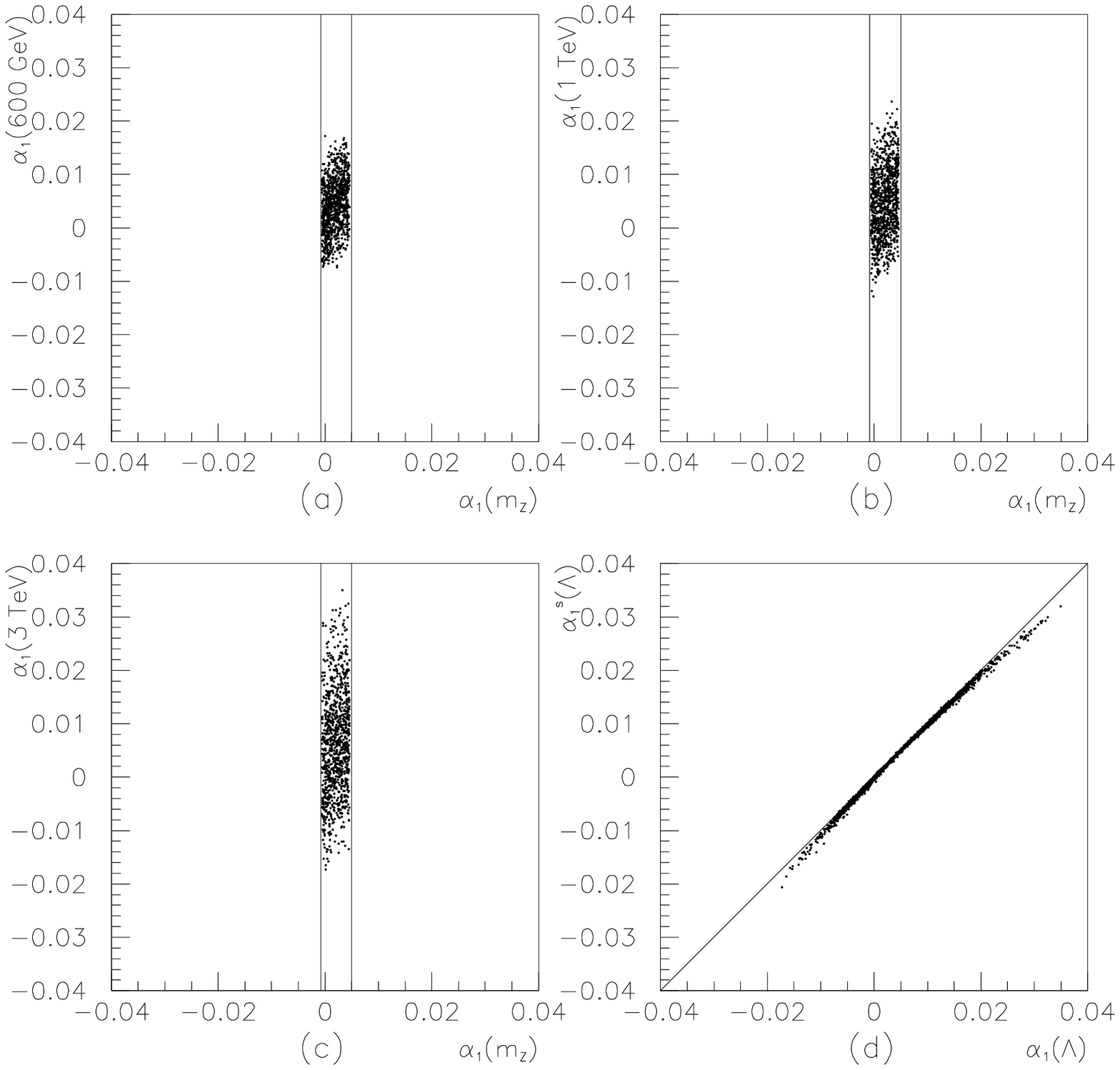,width=\textwidth}
\caption{The running behavior of anomalous coupling $\al_1$,
(a) is devoted to case (1) where $\Lambda=600$ GeV; 
(b) is devoted to case (2) where $\Lambda=1 $ TeV; 
(c) is for case (3) where $\Lambda=3$ TeV.
For these three figures, x axises are the 
value of $\al_1(m_Z)$; y axises are the
value of $\al_1(600 GeV)$, $\al_1(1 TeV)$,
and $\al_1(3 TeV)$, respectively.
Fig. (d) is devoted to compare the value of $\al_1(\Lambda)$ and
$\al_1^s(\Lambda)$ by solving the complete version and simplified 
version of RGEs, respectively; the line $y=x$ is depicted for the sake
of contrast.}
\label{fig4}
\end{figure}

\begin{figure}
\epsfig{file=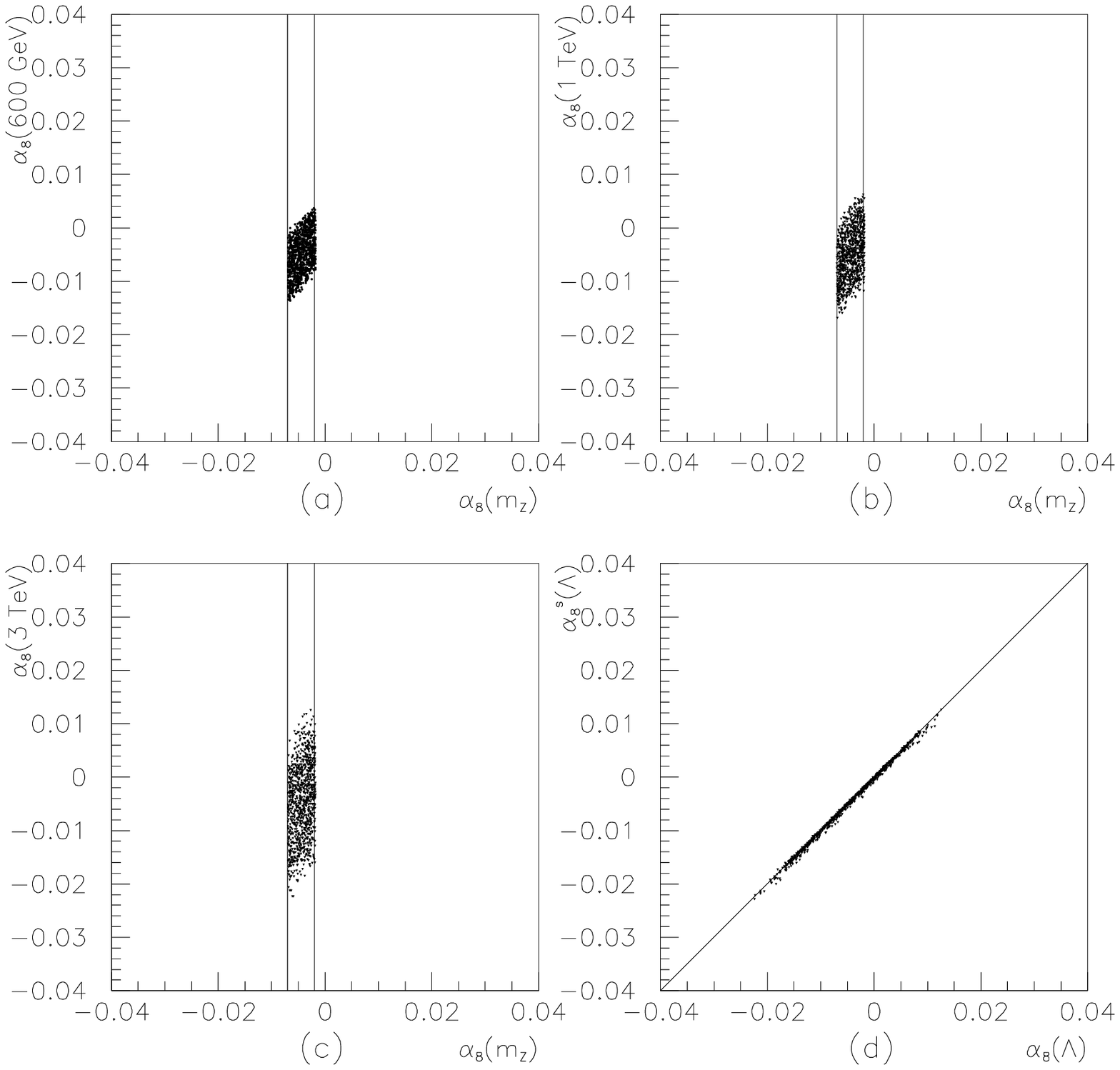,width=\textwidth}
\caption{The running behavior of anomalous coupling $\al_8$,
(a) is devoted to case (1) where $\Lambda=600$ GeV; 
(b) is devoted to case (2) where $\Lambda=1 $ TeV; 
(c) is for case (3) where $\Lambda=3$ TeV.
For these three figures, x axises are the 
value of $\al_8(m_Z)$; y axises are the
value of $\al_8(600 GeV)$, $\al_8(1 TeV)$,
and $\al_8(3 TeV)$, respectively.
Fig. (d) is devoted to compare the value of $\al_8(\Lambda)$ and
$\al_8^s(\Lambda)$ by solving the complete version and simplified 
version of RGEs, respectively; the line $y=x$ is depicted for the sake
of contrast.}
\label{fig5}
\end{figure}

\begin{figure}
\epsfig{file=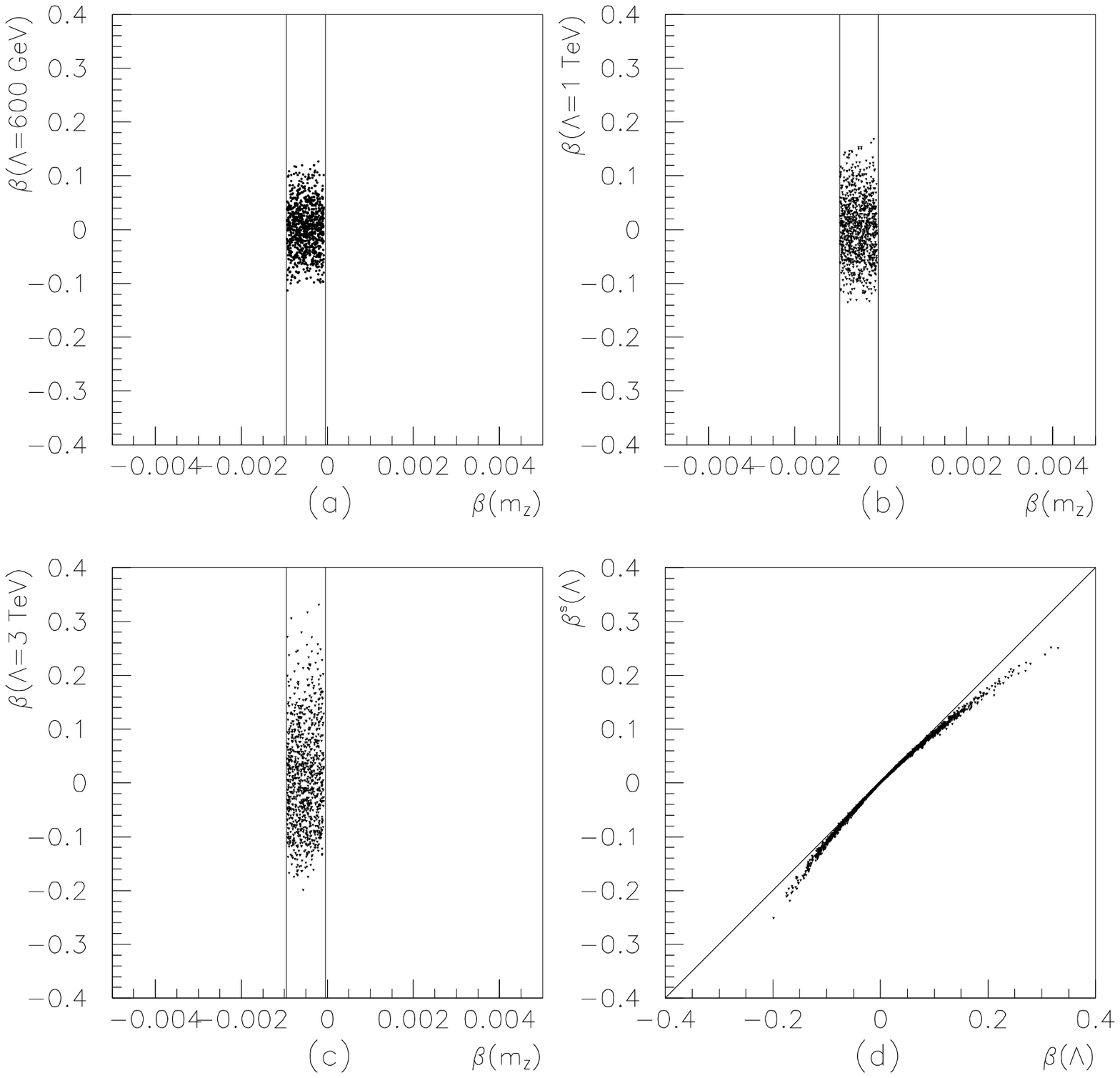,width=\textwidth}
\caption{The running behavior of anomalous coupling $\beta$,
(a) is devoted to case (1) where $\Lambda=600$ GeV; 
(b) is devoted to case (2) where $\Lambda=1 $ TeV; 
(c) is for case (3) where $\Lambda=3$ TeV.
For these three figures, x axises are the 
value of $\beta(m_Z)$; y axises are the
value of $\beta(600 GeV)$, $\beta(1 TeV)$,
and $\beta(3 TeV)$, respectively.
Fig. (d) is devoted to compare the value of $\beta(\Lambda)$ and
$\beta^s(\Lambda)$ by solving the complete version and simplified 
version of RGEs, respectively; the line $y=x$ is depicted for the sake
of contrast.}
\label{fig6}
\end{figure}


\begin{thebibliography}{10}
\bibitem{wein}
	S.Weinberg, Physica {\bf A96} (1979) 327;
	H. Leutwyler, Ann. Phys. NY {\bf  235} (1994) 165.
\bibitem{georgi}
	H. Georgi, Annu. Rev. Nucl. Part. Sci. {\bf 43} (1993) 209.
\bibitem{eff}
	A. Pich,  Published in *Les Houches 1997, 
	Probing the standard model of particle interactions, Pt. 2* 949-1049 
	hep-ph/9806303; J. Wudka, Int. J. Mod. Phys. {\bf A9} (1994) 2301.
\bibitem{long}
	A.C. Longhitano, Phys. Rev. {\bf D22} (1980) 1166; Nucl. Phys. {\bf B188} (1981) 118.
\bibitem{bernard}
	T. Appelquist, C. Bernard, Phys. Rev. {\bf D22} (1980) 200;
	{\it ibid} {\bf D 23} (1981) 425.
\bibitem{appel}
	T. Appelquist, and G. H. Wu, Phys. Rev. {\bf D48} (1993) 3235.
\bibitem{ewcl-fit}
	C.P. Burgess, S. Godfrey, H. Konig, D. London, I. Maksymyk, Phys. Rev. {\bf D49} (1994) 6115.
\bibitem{ewet}
	M.S. Chanowitz, and M.K. Gaillard, Nucl. Phys. {\bf B261} (1985) 379;
	G.J. Gounaris, R. K\"ogerler, and H. Neufeld,
	Phys. Rev. {\bf D34} (1986) 3257;
	H. Veltman, Phys. Rev. {\bf D41} (1990) 2294;
	J. Bagger and C. Schmidt, Phys. Rev. {\bf D41} (1990) 264;
	W. Kilgore, Phys. Lett. {\bf B294} (1992) 257;
	H. J. He, Y. P. Kuang, and X. Li, Phys. Rev. Lett {\bf 69} (1992) 2619;
	Phys. Rev. D 49 (1994) 4842; Phys. Lett. {\bf B329} (1994) 278;
	H.J. He and W.B. Kilgore, Phys. Rev. {\bf D55} (1997) 1515.
\bibitem{scattering}
	M. S. Chanowitz, hep-ph/9812215;
	T. L. Barklow, {\it et al}, hep-ph/9704217.
\bibitem{higgsless}
	 C. Csaki, C. Grojean, H. Murayama, 
	L. Pilo, J. Terning, Phys. Rev. {\bf D69} (2004) 055006.
\bibitem{phen-ewcl}
	K. Hagiwara, T. Hatsukano, S. Ishihara, R. Szalapski, Nucl. Phys. {\bf B496} (1997) 66;
	K. Hagiwara, D. Haidt, S. Matsumoto, Eur. Phys. J. {\bf C2} (1998) 95;
	A. Dobado, J.R. Pelaez, Nucl. Phys. {\bf B425} (1994) 110; Erratum-{\it ibid.} {\bf B434} (1995) 475;
	A. Dobado, J. R. Pelaez, Phys. Lett. {\bf B329} (1994) 469, Addendum-{\it ibid.} {\bf B335} (1994) 554;
	A. Dobado, M.J. Herrero, J.R. Pelaez, E. Ruiz Morales, M.T. Urdiales,
	Phys. Lett. {\bf B352} (1995) 400;
	A. Dobado, M.J. Herrero, J.R. Pelaez, E. Ruiz Morales,
	Phys. Rev. {\bf D62} (2000) 055011;
	A. Dobado, J.R. Pelaez, M.T. Urdiales,
	Phys. Rev. {\bf D56} (1997) 7133;
	J.R. Pelaez, Phys. Rev. {\bf D55} (1997) 4193;
	A. Dobado, Maria J. Herrero, Phys. Lett. {\bf B228} (1989) 495;
	H. J. He, Y. P. Kuang, and C. P. Yuan, Phys. Lett. B 382 (1996) 149;
	A. Dobado, M. J. Herrero, Phys. Lett. {\bf B233} (1989) 505;
	A. Dobado, M. J. Herrero, Tran N. Truong, Phys. Lett. B235 (1990) 129;
	A. Dobado, M. J. Herrero, Juan Terron, Z. Phys. {\bf C50} (1991) 205;
	A. Dobado, M.T. Urdiales, Z. Phys. {\bf C71} (1996) 659;
	R. S. Chivukula, {\it et al.} Annu. Rev. Nucl. Part. Sci. {\bf 45} (1995) 255;
	H.J. He, Y.P. Kuang and C.P. Yuan, in Physics at TeV Energy Scale
	(CCAST-WL Workshop Series: Vol. 72), edited by Y.P. Kuang,
	July 15-26, 1996, CCAST, Beijing, China, pp. 119-234;
	Y.P. Kuang, {\it Lectures at the 2000 summer school on Particle
	Physics and Nuclear Physics}, TUHEP-TH-00115, {\it Electroweak theory II: Electroweak Symmetry Breaking and
	New Physics}, and references therein;
	J. Ellison and J. Wudka, Annu. Rev. Nucl. Part. Sci. {\bf 48} (1998) 33;
	J.M. Butterworth, B.E. Cox, J.R. Forshaw, Phys. Rev. {\bf D65} (2002) 096014;
	The ALEPHI collaboration, Eur. Phys. J. {\bf C21} (2001) 423 [hep-ex/0104034].
\bibitem{review}
	S. Haywood, {\it et. al.}, hep-ph/0003275; W. Kilian, hep-ph/0303015.
\bibitem{manahar}
	A. Manohar, and H. Georgi, Nucl. Phys. {\bf B234} (1984) 189.
\bibitem{smewfit}
	Review of Particle Properties, Phys. Lett. {\bf B592} (2004) 1. 
\bibitem{bagger}
	J.A. Bagger, A. F. Falk, and M. Swartz, Phys. Rev. Lett. {\bf 84} (2000) 1385.
\bibitem{lepexp}
	M.E. Peskin, and J. D. Wells, Phys. Rev. {\bf D 64} (2001) 093003;
\bibitem{lep}
	The L3 Collaboration, hep-ex/0407012.
\bibitem{qcdcl}
	J. Gasser, and H. Leutwyler, Nucl. Phys. {\bf B250} (1985) 465;
	Ann. Phys. 158 (1984) 142.
\bibitem{nonstandard}
	R. S. Chivukula and V. Koulovassilopoulos, Phys. Lett. B 309 (1993) 371;
	D. Kominis, V. Koulovassilopoulos, Phys. Rev. {\bf D52} (1995) 2737;
	T. Han, H.J. He, and C.P. Yuan, hep-ph/9711429.
\bibitem{strong}
	S. Weinberg, Phys. Rev. {\bf D13} (1976) 974; 
	{\it ibid} {\bf D19} (1979) 1277;
	L. Susskind, Phys. Rev. {\bf D20} (1979) 2619.
\bibitem{ntm}
	M. E. Peskin and T. Takeuchi, Phys. Rev. Lett. {\bf 65} (1990) 964;
	Phys. Rev. {\bf D46} (1992) 381.
\bibitem{su2}
	Q. S. Yan and D. S. Du, Phys. Rev. {\bf D69} (2004) 085006;
	S. Dutta, K. Hagiwara, Q. S. Yan, KEK-TH-962, hep-ph/0406090.
\bibitem{div}
	M. Suzuki, Phys. Lett. {\bf B153} (1985), 289;
	M. Kuroda, F.M. Renard and D. Schildknecht, Phys. Lett. {\bf B183} (1987) 366;
	H. Neufeld, J.D. Stroughair and D. Schildknecht, Phys. Lett. {\bf B198} (1987) 563;
	J.A. Grifols, S. Peris and J. Sola, Phys. Lett. {\bf B197} (1987) 437;
	Int. J. Mod. Phys. {\bf A3} (1988) 569;
	J.J. van der Bij, Phys. Lett. {\bf B296} (1992) 239;
	C.P. Burgess and D. London,
	Phys. Rev. Lett. {\bf 69} (1992) 3428; Phys. Rev. {\bf D48} (1993) 4337;
	M.B. Einhorn and J. Wudka,
	{\em Anomalous Vector Boson Couplings---Fact \& Fiction},
	Michigan University preprint UM-TH-92-25 (1992);
	P. Hernandez and F.J. Vegas,
	Phys. Lett. {\bf B307} (1993) 116, hep-ph/9212229.
\bibitem{hagiwara}
	K.~Hagiwara, S.~Ishihara, R.~Szalapski and D.~Zeppenfeld,
	Phys. Rev. {\bf D48} (1993) 2182.
\bibitem{bijj}
	J. J. van der Bij and Boris Kastening, Phys. Rev. {\bf D57} (1998) 2903.
\bibitem{burgess}
	C. P. Burgess, S. Godfrey, H. Konig, D. London, I. Maksymyk, Phys. Rev. {\bf D50} (1994) 7011.
\bibitem{wud}
	G. S$\acute{a}$nchez-Col$\acute{o}$n and J. Wudka, Phys. Lett. {\bf B432} (1998) 383.
\bibitem{path}
	C. Grosse-Knetter and R. K\"ogerler, Phys. Rev. {\bf D48} (1993) 2865;
	S. Dittmaier and C. Grosse-Knetter, Phys. Rev. {\bf D52} (1995) 7276; 
	Nucl. Phys. {\bf B459} (1996) 497.
\bibitem{stuckelberg}
	E. C. G. Stueckelberg., Helv. Phys. Acta {\bf 11} (1938) 299; {\bf 30} (1956) 209;
	T. Kunimasa and T. Goto, Prog. Theor. Phys. {\bf 37} (1967) 524.
\bibitem{bfm}
	B.S. DeWitt, Phys. Rev. {\bf 162} (1967) 1195; {\it ibid.} {\bf 162} (1967)1239;
	't Hooft, Nucl. Phys. {\bf 62} (1973) 444;
	H. Kluber-Stern and J. B. Zuber, Phys. Rev. {\bf D12} (1975) 482;
	M. L\"uscher, Nucl. Phys. {\bf B142} (1982) 359;
	L. F. Abbot, Nucl. Phys. {\bf B185} (1981) 189;
	S. Ichinose and M. Omote, Nucl. Phys. {\bf B203} (1982) 221;
	C. Lee, Nucl. Phys. {\bf B207} (1982) 157;
	I. Jack and H. Osborn, Nucl. Phys. {\bf B207} (1982) 474; {\it ibid}
	{\bf B249} (1985) 472.		
\bibitem{hk}
	Schwinger, Phys. Rev. {\bf 82} (1951) 664;
        R. D. Ball, Phys. Rep. {\bf 182} (1989) 1;
	I. G. Avramidi, {\it Lecture Notes in Physics: N.s. M. Monogrph; 64}
	{\bf Heat Kernel and Quantum Gravity}, Springer (Berlin), 2000;
	D.V. Vassilevich, Phys. Rep. {\bf 388} (2003) 279.
\bibitem{sdem}
	B. S. Dewitt, Phys. Rept. {\bf 19 C} (1975) 295;
	L.S. Brown, Phys. Rev. {\bf D15} (1977) 1469;
	A. Nyffeler, and A. Schenk, Annals Phys. {\bf 241} (1995) 301.
\bibitem{longppr}
	Q.-S. Yan and D.-S. Du, in preparation.
\bibitem{knetter}
	C. Grosse-Knetter, Phys. Rev. {\bf D49} (1994) 6709.
\bibitem{ewclrge}
	 Q.-S. Yan and D.-S. Du, Prepared for 8th Accelerator and Particle Physics 
	Institute (APPI 2003), Appi, Iwate, Japan, 25-28 Feb 2003.
	Published in *Appi 2003, Accelerator and particle physics* 108-125;
	Q.-S. Yan and D.-S. Du, hep-ph/0212367.
\end{thebibliography}
\end{document}